\documentclass{article}
\usepackage{graphicx}
\usepackage{multirow}
\usepackage{longtable,booktabs}
\usepackage[centertags]{amsmath}
\usepackage{bm}
\usepackage{authblk}
\usepackage{amsthm,natbib}
\RequirePackage[colorlinks]{hyperref}
\RequirePackage{natbib}

\newcommand{\E}{\mbox{E}}

\newcommand{\y}{\mathbf{y}}
\newcommand{\ub}{\mathbf{u}}

\newcommand{\uno}{\mathbf{1}}

\DeclareGraphicsExtensions{.pdf,.eps,.png,.jpg}

\title{Multilevel Models with Stochastic Volatility for Repeated Cross-Sections: an Application to tribal Art Prices}
\author{Silvia Cagnone \thanks{silvia.cagnone@unibo.it}}
\author{Simone Giannerini \thanks{simone.giannerini@unibo.it}}
\author{Lucia Modugno \thanks{lucia.modugno@unibo.it} }
\affil{Dipartimento di Scienze Statistiche, University of Bologna, Italy}
\date{}
\begin{document}
\maketitle

\begin{abstract}
In this paper we introduce a multilevel specification with stochastic volatility for repeated cross-sectional data. Modelling the time dynamics in repeated cross sections requires a suitable adaptation of the multilevel framework where the individuals/items are modelled at the first level whereas the time component appears at the second level. We perform maximum likelihood estimation by means of a nonlinear state space approach combined with Gauss-Legendre quadrature methods to approximate the likelihood function. We apply the model to the first database of tribal art items sold in the most important auction houses worldwide. The model allows to account properly for the heteroscedastic and autocorrelated volatility observed and has superior forecasting performance. Also, it provides valuable information on market trends and on predictability of prices that can be used by art markets stakeholders.
\end{abstract}


\section{Introduction}\label{sec:intro}
The investigation of the relationship between the art market and financial markets has important implications for institutions as well as for auction houses, art merchants and individuals. In fact, also due to the recent financial crisis, there has been a sharp increase in the so called alternative investments that comprise funds specialising in art. These appear to offer a highly beneficial diversification strategy with a complex correlation with traditional assets. Hence, the study of the features of this new asset class is important and cannot disregard the differences with respect to traditional stocks. For instance, art items are exchanged a few times and the transaction costs are considerable. Moreover, there is the so called \emph{aesthetic dividend} which plays a crucial role \citep[see e.g.][and references therein]{Goetzmann, Candela13}.
\par
The study of price determination and price indexes for art items is fundamental to auction houses and art merchants. Besides estimating the price of individual items, price indexes can be used to understand market trends, to assess the main social and economic factors that influence the art market, to understand whether investing in art would diversify risk in a long-term investment portfolio. For art objects, the traditional view of a long run price related to the cost of production does not hold anymore. Some authors argue that the art market is inherently unpredictable since it is dictated by collectors' manias \cite{Bau86}. However, there is an ever growing consensus, backed by empirical evidence, on the idea that ``price fundamentals'' can be objectively identified.
The hedonic regression, also known as the \emph{grey painting method}, is one of the most used approaches for modelling art prices. It was first proposed in \cite{Rosen} and further investigated and applied in~\cite{Chanel, Ginsburgh, Agnello, Chanel2, Locatelli, Collins}. According to this method, the price of an artwork item depends both on market trends and on a set of characteristics of the item itself. Such dependence is modelled through a fixed effect regression and the estimated regression coefficients can be interpreted as the price of each feature, the so-called \emph{shadow price}. Hence, it is possible to predict the price of a given object by summing the prices of its features. Also, a time-dependent intercept can represent the value of the \emph{grey painting} in that period, that is, the value of an artwork created by a standard artist, through standard techniques, with standard dimensions, etc.~\citep{Locatelli}. Eventually, the price index is built from the prices of the \emph{grey painting} in different periods.
\par
Despite its potential, the hedonic regression model has several shortcomings. First, as also remarked in \citet{Goet15} only a small fraction of the great variability of the price dynamics is explained. Second, most of the features are categorical so that the regression equation contains many dummy variables and the resulting models are not parsimonious. Most importantly, the time dynamics is not modelled directly but through dummy variables so that it is not possible to use the model to forecast the prices. Moreover, since it is practically impossible to follow the selling price of each artwork item over time, the available datasets have a structure of a repeated cross-section where at each time point a new sample is observed.
\par
In order to overcome the above-mentioned issues, a multilevel approach can be used \citep{Goldstein2010, Skrondal_book}. The treatment of repeated cross-sectional data requires the extension of the classical multilevel model by considering individual heterogeneity within time at the first level, and the variability over time at the second level. This specification has been adopted for the first time by \cite{Diprete} and \cite{Browne} in the frequentist and Bayesian frameworks, respectively. Motivated by the construction of a price index for auctioned items of tribal artworks, \citet{Modugno15} extended such multilevel specification as to add an autoregressive components at the second level. They found a considerable improvement over classical models in terms of prediction and forecasting but the assumption of normality of level-1 residuals was violated, probably due to the presence of heteroscedasticity and kurtosis. They devised an ad hoc solution for deriving robust standard errors through the wild bootstrap scheme for multilevel models introduced in \cite{Mod15}. Similar findings concerning the non normality and the heteroscedasticity were reported for other specifications for the art market in \cite{Bocart11} and \cite{Hod04}. In \cite{Bocart11} the problem  was addressed by estimating a semiparametric time-varying volatility and Student's t error with skewness, whereas \cite{Hod04} did not assume any parametric form for the disturbances but retained the assumption of serial independence of random effects. Also, \cite{Bocart13} modeled the volatility of price indexes by means of a smooth function of time as a component of an unbalanced panel model with AR(1) time effects. They implemented a linear Gaussian state-space representation and estimated it through maximum likelihood combined with a Kalman filter, and, as above, they found a violation of the normality assumption. The evidence reported in literature indicates that the volatility of prices plays a pre-eminent role; assuming it constant is not realistic and might cause estimation problems.
\par
In this paper we extend the model proposed in \citet{Modugno15} by including a stochastic volatility component at the second level by means of a nonlinear state space approach. The specification is motivated by the analysis of the first world database of tribal art prices. This allows to account properly for the heteroscedastic and autocorrelated volatility of level-1 error terms and brings in several advantages. 
Stochastic volatility models (SV) are based on the assumption that the conditional variance of the observed variable depends on a latent variable that captures the flow of information arriving from the market. Similar to ARCH-type models for financial time series, SV models allow to account properly for fat tailed distributions, white-noise-type dependence, high and lag-decreasing autocorrelations of squared observations.  We opt for a stochastic volatility component since ARCH-type models assume that the volatility is affected  by past information through a deterministic function. Such a specification is not viable for repeated cross-sections.
\par
Model estimation is performed through maximum likelihood via a non-linear Gaussian filtering process in the spirit of \cite{Kitagawa} and \cite{Tani95}. The task poses several computational challenges related to the presence of time-varying latent variables that must be integrated out from the likelihood function so that there is no analytical solution. To this aim, \cite{Fridman} proposed a non-linear Kalman filter algorithm by expressing the likelihood function as a nested sequence of one-dimensional integrals approximated by the Gauss Legendre numerical quadrature. \cite{Bartolucci} extended this approach by computing analytical first and second derivatives of the approximated likelihood. They applied a rectangular quadrature to approximate the integrals. More recently, \cite{BaCa16}  approximated such integrals by using an Adaptive Gauss Hermite quadrature method. Here, we extend the procedure discussed in \cite{Fridman} by using Gauss-Legendre quadrature methods to approximate the integrals involved in the likelihood.

\section{The first database of tribal art prices}\label{sec:data}
 The first database of tribal art prices was created in 2006 from the joint agreement of the department of Economics of the University of the Italian Switzerland, the Museum of the Extra-European cultures in Lugano, the Museo degli Sguardi in Rimini, and the Faculty of Economics of the University of Bologna, campus of Rimini. For each artwork item, there are 37 variables recorded from the catalogues released before the auctions. The variables include physical, historical and market characteristics. Some of these are shown in Table~\ref{tab:data1} and most of them are categorical.
 \begin{table}
  \footnotesize
     \caption{Subset of variables classified by type: Physical, Historical and Market.}\label{tab:data1}
    \centering
    \begin{tabular}{l l l}
    \toprule
    & \textit{Variable}&\textit{Categories}\\
      \hline
\multirow{12}{*}{\emph{Physical}}
    & Type of object& Furniture, Sticks, Masks,\\
    &               & Religious objects, Ornaments,\\
    &               & Sculptures, Musical instruments,\\
    &               & Tools, Clothing, Textiles,\\
    &               & Weapons, Jewels \\
    &  Material     & Ivory, Vegetable fibre, Wood,\\
    &               & Metal, Gold, Stone,\\
    &               & Precious stone, Terracotta, ceramic,\\
    &               & Silver, Textile and hides\\
    &               & Seashell, Bone, horn, Not indicated  \\
    &  Patina       & Not indicated, Pejorative,\\
    &               & Present, Appreciative \\
      \hline
\multirow{15}{*}{\emph{Hystorical}}
    &  Continent                       & Africa,  America\\
    &                                  & Eurasia, Oceania\\
    &  Illustration on the catalogue   & Absent, Black/white, Coloured,\\
    &  Illustration width              & Absent, Miscellaneous, \\
    &                                  & Quarter page, Half page,\\
    &                                  & Full page, More than one,\\
    &                                  & Cover\\
    & Description                      & Absent, Short visual, Visual,\\
    &                                  &    Broad visual, Critical,       Broad critical.\\
    & Specialized bibliography         & Yes, No\\
    & Comparative bibliography         & Yes, No\\
    & Exhibition                       & Yes, No\\
    &  Historicization                 & Absent, Museum certification,\\
    &                                  & Relevant museum certification,\\
    &                                  & Simple certification   \\
      \hline
\multirow{2}{*}{\emph{Market}}
    &  Venue                           & New York, Paris,\\
    &  Auction house                   & Sotheby's,            Christie's\\
   \bottomrule
    \end{tabular}
\end{table}
After the auction, the information on the selling price is added to the record.
 \begin{figure}
    \centering
   \includegraphics[keepaspectratio=true,width=11cm]{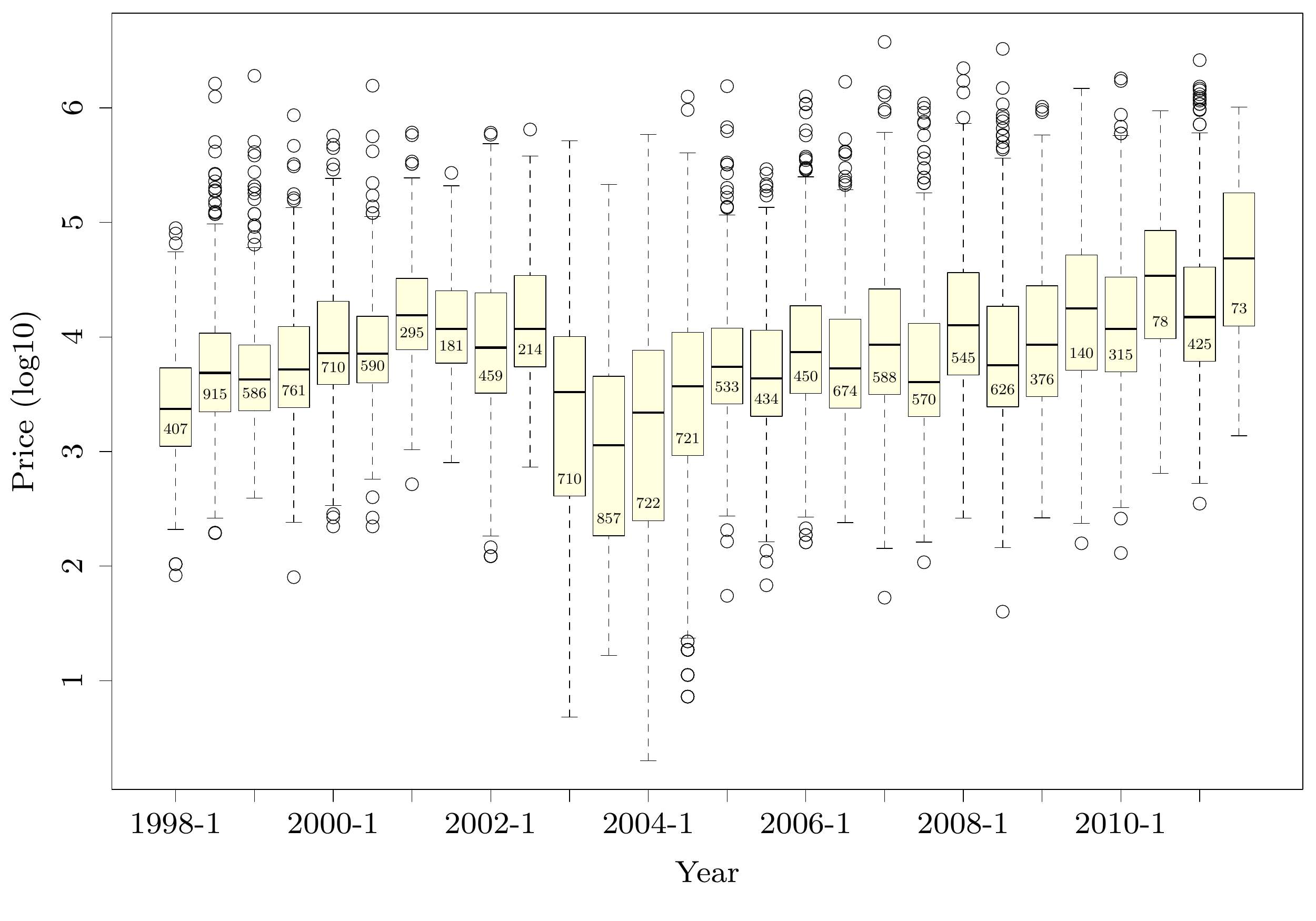}
    \caption{Boxplots of prices in logarithmic scale (base 10) of the tribal art market by semester. The amount of items sold in a given semester is reported inside the boxes.}\label{fig:series}
\end{figure}

Figure \ref{fig:series} shows the boxplots of logged prices aggregated by semester; the number of items sold in each semester is reported inside the boxes. The structure of the dataset emerges clearly from the graph: in every semester a different group of artworks is sold; e.g. 407 items were auctioned in 1998-1, 915 objects different from the first set were sold in 1998-2, and so on. Hence, tribal art data has a structure like that of repeated cross-sectional surveys and the medians (black lines) give an idea of the trend of prices over time. In particular, note the consistent reduction in the number of auctioned items starting from 2009. Despite this, the overall turnover did not drop since the average price rose. This might indicate the adoption by market agents of hedging strategies against the economic crisis. Overall, we have $T=28$ semesters, and $n_t$, the number of items sold in the semester $t$, varies between 73 (2011-2) and 915 (1998-2); the total sample size of sold items  (\mbox{$n=\sum_{t=1}^T n_t$}) is 13955. It is convenient to aggregate the data in semesters rather than auction dates since auction dates are not equally spaced in time and this feature is important for modelling time dependence. Also, auctions are organized in two sessions, one in the winter and one in the summer, and each session contains two to four auctions quite close in time. The aggregation in semesters respects naturally this organization so that the results are meaningful from the Economics point of view and this source of heterogeneity can be explained.


\section{The model}\label{application}
In Section~\ref{sec:M1} we briefly review the model proposed in \cite{Modugno15} while in Section~\ref{sec:ML} we extend it by introducing the mixed effects model with stochastic volatility.

\subsection{A Multilevel model with autoregressive random effects}\label{sec:M1}
Let $y_{it}$ be the observed price for item $i=1,\ldots,n_t$ at time-point $t=1,\ldots, T$ and let $\mathbf{x}_{it}$ be a corresponding column vector of $k$ covariates. Since tribal art data can be thought to have a two-level structure where items represent level-1 units, and time points represent level-2 units, we consider the following random intercept model:
\begin{equation*}\label{eq:ARE1}
\log_{10}(y_{it})= \beta_{0}+ u_{t}+\mathbf{x}'_{it}\bm{\beta}+
\epsilon_{it},\quad \epsilon_{it}|\mathbf{x}_{t} \sim
\mathrm{NID}(0,\sigma^2)
\end{equation*}
where $u_{t}$ are time-specific random intercepts that account for the unobserved heterogeneity between items within each time point, $\bm{\beta}$ is a vector of fixed slopes and $\beta_0$ is the overall mean. In repeated cross-sectional data, $y_{it}$ and $y_{i(t+1)}$ are not the price of the same item $i$ observed at successive time points since the two objects are physically different. Conditionally on the vector of covariates $\mathbf{x}_{t}$, level-1 errors $\epsilon_{it}$ (the error term for a given individual at a given time point) follows a normal distribution with constant variance. In other words, the art market is assumed to have a constant volatility over time. Note that $\epsilon_{it}$ is conditioned on the vector of the covariates $\mathbf{x}_{t}$ for all the individuals, that is, we assume strict exogeneity on the explanatory variables ~\citep{Woold:02}. Differently from panel data, in repeated cross-sectional data the strict exogeneity assumption implies that, for each item, the covariates are uncorrelated with the error terms.
\par
The dynamics of $u_{t}$ can be modelled at the second level by extending the above multilevel models as follows
\begin{eqnarray*}\label{eq:ARE2}
u_{t}=\rho u_{t-1}+\eta_t, \qquad \eta_t|\mathbf{x}_t \sim \mathrm{NID}(0,\sigma^2_{\eta}),
\end{eqnarray*}
where $\eta_t \bot u_s$  and $\eta_t \bot \epsilon_{it}$ for all $s<t$ and for all $i$. In this specification, the random effects follow an autoregressive process of order 1 with $|\rho| < 1$, that guarantees stationarity. We denote this model as ARE (Autoregressive Random Effects). \citet{Modugno15} introduced a full maximum likelihood estimation method via the EM algorithm to fit the ARE model to the tribal art data. The ARE model improves considerably over classical models in terms of prediction and forecasting. However, as it will be shown in the application Section, the assumption of normality of level-1 residuals is violated, probably due to the presence of heteroscedasticity and kurtosis. As mentioned above, the assumption of a constant volatility of prices of art assets is not realistic and might cause severe inference problems. In the following we extend the ARE model by including a stochastic volatility component that accounts properly for the heteroscedastic and autocorrelated volatility of the level-1 error process.

\subsection{A Multilevel model with autoregressive random effects and stochastic volatility}\label{sec:ML}
We include a stochastic volatility component at level-1 of the ARE model as follows:
\begin{eqnarray} \label{SS1}
\log_{10}(y_{it})&=&\beta_0+u_t+ \mathbf{x}'_{it}\bm{\beta}+\exp{(h_t/2)}\epsilon_{it} \\
u_t&=&\rho u_{t-1}+\eta_t\\ \label{SS2}
h_t&=&\alpha + \delta h_{t-1} + \sigma_{\nu}\nu_t, \label{SS3}
\end{eqnarray}
for $i=1, \ldots,n_t$ and $t=1,\ldots,T$. As before, $u_{t}$ is the time dependent random effect whereas $h_t$ is the latent variable that represents the volatility component at time $t$. Both $u_t$ and $h_t$ follow a stationary autoregressive process, so that $|\rho|<1$ and $|\delta|<1$. We take the corresponding unconditional distributions as initial conditions for the two processes, namely $h_1 \sim N\big( \alpha/(1-\delta),\sigma^2_{\nu}/ (1-\delta^2)\big)$ and $u_1 \sim N\big(0,\sigma^2_{\eta}/(1-\rho^2)\big)$.
Moreover, we assume that $\epsilon_{it}$, $\eta_t$ and $\nu_t$  are mutually independent, with $\epsilon_{it} \sim \mathrm{NID}(0,1)$, $\eta_t\sim \mathrm{NID}(0,\sigma^2_{\eta})$ , and $\nu_t\sim \mathrm{NID}(0,1)$ respectively, where, for simplicity of notation, we omit the conditioning upon the covariates $\mathbf{x}_{t}$.
Under these assumptions, the conditional densities result
\begin{eqnarray*}
y_{it}|u_t,h_t &\sim& \mathrm{NID}(\beta_0+u_t+\mathbf{x}'_{it}\bm{\beta},\exp{(h_t)})\\
u_t|u_{t-1}&\sim& \mathrm{NID}(\rho u_{t-1},\sigma^2_{\eta})\\
h_t|h_{t-1} &\sim& \mathrm{NID}(\alpha + \delta h_{t-1},\sigma^2_{\nu}).
\end{eqnarray*}
 We call the new specification SVARE (Stochastic Volatility and Autoregressive Random Effects) model. The system (\ref{SS1})-(\ref{SS3}) is a nonlinear state space representation. Hence, model estimation can be performed by using maximum likelihood via a non-Gaussian filtering process and poses several non-trivial challenges that we describe and address in the following section.

\section{Model estimation}
\subsection{The likelihood function}\label{sec:lik}
The maximum likelihood estimation is based on the maximization of the following likelihood function
\begin{equation} \label{lik1}
\begin{split}
L(\bm{\theta}|\y)=&\int_{\mathbf{h}}\int_{\ub} f(\y|\ub,\mathbf{h})f(\ub,\mathbf{h})\mbox{d}\ub \mbox{d}\mathbf{h} = \\
=&\int_{-\infty}^{+\infty}\ldots\int_{-\infty}^{+\infty}\left[\prod_{t=1}^{T}f(\y_{t}|u_t,h_t)f(u_t|u_{t-1})f(h_t|h_{t-1})\right] \mbox{d}u_T\ldots \mbox{d}u_1\mbox{d}h_T\ldots \mbox{d}h_1\\
\end{split}
\end{equation}
where $\bm{\theta}=\{\beta_0,\bm{\beta}',\rho,\sigma_{\eta},\alpha,\delta,\sigma_{\nu}\}$ is the vector of parameters, $f(\y_{t}|u_t,h_t)=\prod_{i=1}^{n_t}f(y_{it}|u_t,h_t)$ for $t=1,\ldots,T$, $f(u_1|u_{0})=f(u_1)$ and $f(h_1|h_{0})=f(h_1)$. \par
The computation of  $L(\bm{\theta}|\y)$ requires solving a $2T$-dimensional integral which is computationally unfeasible. We address the issue by applying an iterated numerical integration procedure introduced by \cite{Kitagawa} for non-Gaussian filtering problems. The procedure is based upon rephrasing the likelihood (\ref{lik1}) as
\begin{equation} \label{lik2}
\begin{split}
L(\bm{\theta}|\y)&=\int_{-\infty}^{+\infty}\int_{-\infty}^{+\infty}f(\y_{1}|u_1,h_1)f(u_1)f(h_1)\int_{-\infty}^{+\infty}\int_{-\infty}^{+\infty}f(\y_{2}|u_2,h_2)f(u_2)f(h_2)\ldots\\
&\ldots\int_{-\infty}^{+\infty}\int_{-\infty}^{+\infty}f(\y_{T}|u_T,h_T)f(u_T)f(h_T)\mbox{d}u_T\mbox{d}h_T \ldots \mbox{d}u_2\mbox{d}h_2 \mbox{d}u_1\mbox{d}h_1.\\
\end{split}
\end{equation}
 The resulting bivariate integrals can be approximated by using numerical quadrature techniques. The most common quadrature techniques for stochastic volatility models are the rectangular quadrature \citep{Bartolucci}  and the Gauss-Legendre quadrature rule \citep{Fridman}. We adopt the latter method since in this context we found it superior with respect to other proposals, especially in terms of computational time. The application of the Gauss-Legendre quadrature rule to eq. (\ref{lik2}) produces the following approximated likelihood function
\begin{equation}\label{eq:lik_appr}
\begin{split}
\tilde{L}(\bm{\theta}|\y)=&\bigg(\frac{b-a}{2}\bigg)^T\bigg(\frac{e-d}{2}\bigg)^T \sum_{i_1}^{n_u}w_{ui_1}\sum_{j_1}^{n_h}w_{hj_1}
 f(\y_{1}|u^*_{i_1},h^*_{j_1}) f(u^*_{i_1})f(h^*_{j_1})\\
&\sum_{i_2}^{n_u}w_{ui_2}\sum_{j_2}^{n_h}w_{hi_2}
f(\y_{2}|u^*_{i_2},h^*_{j_2}) f(u^*_{i_2}|u^*_{i_1})f(h^*_{j_2}|h^*_{j_1})\ldots\\
&\ldots \sum_{i_T}^{n_u}w_{ui_T}\sum_{j_T}^{n_h}w_{hi_T}
 f(\y_{T}|u^*_{i_T},h^*_{j_T})f(u^*_{i_T}|u^*_{i_{T-1}})f(h^*_{j_T}|h^*_{j_{T-1}})
\end{split}
\end{equation}
where
 $\{u^*_i\}$, with $i=1,\ldots,n_u$, and $\{h^*_j\}$, with $j=1,\ldots,n_h$, are sets of Gauss-Legendre quadrature points, $w_{ui}$ and $w_{hj}$ are the corresponding weights and $[a,b]$ and $[d,e]$ are finite integration limits which replace the infinite ones for the random effects and the volatility process respectively. The choice of the grids and the number of evaluation points is crucial for the numerical precision. First, as proposed in~\cite{Fridman}, the grids for the two latent processes are centered on $\mu_u=0$ and $\mu_h=\alpha/(1-\delta)$ with a width of $3\sigma_u=3\sigma_{\eta}/(\sqrt{1-\rho^2})$ and $3\sigma_h=3\sigma_{\nu}/(\sqrt{1-\delta^2})$; this allows the grids to cover the support of the unconditional distributions with non negligible mass. Second, the number of quadrature points, $n_u$ and $n_h$, are chosen according to the degree of smoothness of the integrands, i.e. the average distance between two points is less or equal to $\sigma_{\eta}/2$ for the random effect process and $\sigma_{\nu}/2$ for the volatility process. \par

 \subsection{Recursive algorithm}
 In order to facilitate the computation of  the approximated likelihood (\ref{eq:lik_appr})  we use the matrix notation for the  recursive algorithm. Define  the $(n_u \times 1)$ and $(n_h \times 1)$ vectors of the initial probabilities as
$$ \mathbf{f}_{u_1} = (f(u^*_1), f(u^*_2), \ldots,f(u^*_{n_u}))'\quad \mathrm{and} \quad \mathbf{f}_{h_1} = (f(h^*_1), f(h^*_2), \ldots,f(h^*_{n_h}))'$$and the $(n_un_h \times 1)$ vector of the initial conditional probabilities as
$$\mathbf{f}_1 = (f(\y_{1}|u^*_1,h^*_1), f(\y_{1}|u^*_2,h^*_1), \ldots,f(\y_{1}|u^*_{n_u},h^*_1), f(\y_{1}|u^*_1,h^*_2), \ldots, f(\y_{1}|u^*_{n_u},h^*_{n_h}))'$$
In the same way, define the probability transition matrices of the two discretized latent variables as
$$\mathbf{F}_h=\left[ \begin{array}{ccc}f(h^*_1|h^*_1)&\ldots& f(h^*_1|h^*_{n_h})\\
\vdots & \vdots& \vdots\\
f(h^*_{n_h}|h^*_1)&\ldots& f(h^*_{n_h}|h^*_{n_h})\end{array} \right] \quad \mathbf{F}_u=\left[ \begin{array}{ccc}f(u^*_1|u^*_1)&\ldots& f(u^*_1|u^*_{n_u})\\
\vdots & \vdots& \vdots\\
f(u^*_{n_u}|u^*_1)&\ldots& f(u^*_{n_u}|u^*_{n_u})\end{array} \right]$$
and the $(n_un_h \times 1)$ vector of conditional probabilities for $t=2,\ldots,T$ as
$$\mathbf{f}_t =(\mathbf{f}_{t\_1}, \mathbf{f}_{t\_2}\ldots \mathbf{f}_{t\_n_{h}})' \quad \mathrm{with} \quad \mathbf{f}_{t\_j}= (f(\mathbf{y}_{t}|u^*_1,h^*_j), f(\mathbf{y}_{t}|u^*_2,h^*_j),\ldots,f(\mathbf{y}_{t}|u^*_{n_u},h^*_j))'$$
Finally, define the vectors $\mathbf{w}_u=(w_{u1},w_{u2},\ldots,w_{un_u})'$, $\mathbf{w}_h=(w_{h1},w_{h2},\ldots,w_{hn_h})'$ and the matrices $\mathbf{W}_u=(\mathbf{w}_{u},\mathbf{w}_{u},\ldots,\mathbf{w}_{u})$, $\mathbf{W}_h=(\mathbf{w}_{h},\mathbf{w}_{h},\ldots,\mathbf{w}_{h})$ of dimensions $(n_u\times n_u)$ and $(n_h\times n_h)$ respectively.  The approximated marginal likelihood can be computed through the following iterative algorithm

\begin{enumerate}
   \item compute the joint probability at the first time point ($t=1$):
     \begin{equation}\label{eq:st1}
       \mathbf{l}_1=\left(\frac{b-a}{2}\bigg)\bigg(\frac{e-d}{2}\right)\text{diag}(\mathbf{f}_1)((\mathbf{f}_{h_1} \circ\mathbf{w}_h)\otimes (\mathbf{f}_{u_1}\circ \mathbf{w}_u))
     \end{equation}
     where $\circ$ indicates the Hadamard product between two matrices and $\text{diag}(\mathbf{f}_1)$ is the block-diagonal matrix created from the vector $\mathbf{f}_1$. The generic element of $\mathbf{l}_1$ is $l_{1ij}=f(\y_1,u_{t}=u_i^*,h_{t}=h_j^*)$.
   \item  compute the forward intermediate joint probabilities for $t = 2,\ldots,T$:
    \begin{equation}\label{eq:forw} \mathbf{l}_t=\bigg(\frac{b-a}{2}\bigg)\bigg(\frac{e-d}{2}\bigg)\text{diag}(\mathbf{f}_t)((\mathbf{F}_h \circ \mathbf{W}_h)\otimes (\mathbf{F}_u \circ \mathbf{W}_u) ) \mathbf{l}_{t-1},
 \end{equation}
   where $\text{diag}(\mathbf{f}_t)$ is the block-diagonal matrix created from the vector $\mathbf{f}_t$. The generic element of $\mathbf{l}_t$ is $l_{tij}=f(\y_1,\ldots,\y_t,u_{t}=u_i^*,h_{t}=h_j^*)$.
   \item compute the  approximated marginal likelihood (\ref{eq:lik_appr}) as
   \begin{equation}\label{eq:L}\tilde{L}(\boldsymbol{\theta}|\y)=
\uno'\mathbf{l}_T
\end{equation}
 where $\uno$ is a $(n_un_h\times 1)$ vector of ones.
 \end{enumerate}
The maximization of expression~(\ref{eq:L}) is carried out by using a quasi-Newton algorithm in which the gradient and the Hessian matrix are obtained through numerical derivatives. The choice of the starting values for $\bm{\theta}$ requires two separate strategies, one for the parameters of the stochastic volatility component and one for the remaining parameters. As for the former, we adapt to repeated cross-sectional data the method used by \cite{Bartolucci} in a time series context. First, consider the stochastic volatility model for the error terms of the ARE model
\begin{equation}\label{eq:ini}
y_{it}^*=y_{it}-\beta_0 - u_t - \mathbf{x}'_{it}\bm{\beta}=\exp(h_t/2)\epsilon_{it},
\end{equation}
from which, as observed by \cite{Har94}, the following expression can be derived  \[\log y_{it}^{*2}=h_t + \log \epsilon_{it}^2\]
Since $\E(\log \epsilon_{it}^2)=-1.27$ we have that
\[h_t \approx h_t^*= \log \hat{y}_{it}^{*2}+1.27\] where $\hat{y}_{it}$ are obtained as the level-1 residuals of the ARE model. Second, smooth $h_t^*$ through a $k$-term moving average, run a regression of $h_t^*$ on $h_{t-1}^*$ and use the resulting estimates as initial values of the stochastic volatility parameters. As concerns the remaining parameters, the estimates of the fitted ARE model are taken as starting values.
\subsection{Filtering, Smoothing and Prediction}
In order to obtain optimal estimators of the unobserved-state vectors $\mathbf{u}$ and $\mathbf{h}$, we perform filtering and smoothing, which differ in the conditioning information set. Moreover, we derive the one-step-ahead predictors.

\subsubsection{Filtering}
Denoting with $\mathbf{Y}_{t}=(\mathbf{y}_1,\ldots,\mathbf{y}_{t})$, the filtered volatility and the filtered random effects at time $t$ are defined as
\begin{align*}
\hat{h}_t(\mathbf{Y}_t)&=E(h_t|\mathbf{Y}_t)=\int h_t f(h_t|\mathbf{Y}_{t}) dh_t\\
\hat{u}_t(\mathbf{Y}_t)&=E(u_t|\mathbf{Y}_t)=\int u_t f(u_t|\mathbf{Y}_{t}) du_t.
\end{align*}

 The filtered values are computed directly from the quantities obtained in the forward recursive algorithm:
 \begin{align*}
 \hat{h}_t(\mathbf{Y}_t)&= \int h_t\frac{f(h_t,\mathbf{Y}_{t})}{f(\mathbf{Y}_{t})}dh_t=\frac{\int  \int h_t f(\mathbf{Y}_{t},u_{t},h_{t}) du_{t} dh_t}{\int\int f(\mathbf{Y}_{t},u_{t},h_{t}) du_{t}dh_{t}}\simeq  \frac{\sum_j^{n_h} h_j^* \sum_{i}^{n_u} l_{tij}}{\sum_{i}^{n_u}\sum_j^{n_h} l_{tij}}\\
\hat{u}_t(\mathbf{Y}_t)&= \int u_t\frac{f(u_t,\mathbf{Y}_{t})}{f(\mathbf{Y}_{t})}du_t=\frac{\int  \int u_t f(\mathbf{Y}_{t},u_{t},h_{t}) dh_{t} du_t}{\int\int f(\mathbf{Y}_{t},h_{t},u_{t}) du_{t}dh_{t}}\simeq  \frac{\sum_i^{n_u} u_j^* \sum_{j}^{n_h} l_{tij}}{\sum_{i}^{n_u}\sum_j^{n_h} l_{tij}}
\end{align*}

 \subsubsection{Smoothing}\label{sec:smooth}
In the multilevel framework, the random effects are predicted by using either the Best Linear Unbiased Predictor or the Empirical Bayes Predictor. In both cases, the best predictor is the expected value conditioned to the whole observed sample or empirical posterior distribution (for an overview see \cite{Searle} and \cite{Skrondal_book}). In stochastic volatility models the smoothed values are:
 \begin{align}
\hat{h}_t(\mathbf{Y}_T)&=E(h_t|\mathbf{Y}_T)=\int\int h_t\frac{f(\mathbf{Y}_{t},h_{t},u_{t})f(\y_{t+1},\ldots,\y_T|h_{t},u_{t})}{f(\mathbf{Y}_{T})}du_tdh_t \label{ht}\\
\hat{u}_t(\mathbf{Y}_T)&=E(u_t|\mathbf{Y}_T)=\int \int u_t\frac{f(\mathbf{Y}_{t},h_{t},u_{t})f(\y_{t+1},\ldots,\y_T|h_{t},u_{t})}{f(\mathbf{Y}_{T})}dh_tdu_t.\label{ut}
\end{align}
In both expressions, the density $f(\mathbf{Y}_{T})$ is approximated through the marginal likelihood $\tilde{L}(\boldsymbol{\theta}|\y)$ of Eq. (\ref{eq:L}) and $f(\mathbf{Y}_{t},h_{t},u_{t})$ through $l_{tij}$. The intermediate conditional probabilities, $f(\y_{t+1},\ldots,\y_T|h_{t},u_{t})$ can be obtained by using the following backward recursion
 \begin{enumerate}
   \item Define $\mathbf{b}_T=\uno_{n_hn_u}$.
   \item Compute the backward conditional probability at time $t$, with  $t=T-1,\ldots,1$:
   $$
   \mathbf{b}_t=\bigg(\frac{b-a}{2}\bigg)\bigg(\frac{e-d}{2}\bigg)\text{diag}(\mathbf{f}_{t+1})((\mathbf{F}_h \circ \mathbf{W}_h)\otimes (\mathbf{F}_u \circ \mathbf{W}_u) ) \mathbf{b}_{t+1}
   $$
   where the generic element is  $b_{tij} = f(\mathbf{y}_{t+1},\ldots,\y_T|u_{t}=u^*_i,h_{t}=h^*_j)$.
 \end{enumerate}
 With this further recursion, the smoothed values are approximated as
\[
\hat{h}_t(\mathbf{Y}_T)\simeq  \frac{\sum_j^{n_h} h_j^* \sum_{i}^{n_u} l_{tij}b_{tij}}{\tilde{L}(\boldsymbol{\theta}|\y)}\;;
\quad
\hat{u}_t(\mathbf{Y}_T)\simeq  \frac{\sum_i^{n_u} u_i^* \sum_{j}^{n_h} l_{tij}b_{tij}}{\tilde{L}(\boldsymbol{\theta}|\y)}.
\]

\subsubsection{Prediction}
The one-step-ahead predicted values of volatility and random effects are derived through the following approximations:
\begin{align*}\label{eq:h_hat}
\hat{h}_t(Y_{t-1})&=E(h_t|Y_{t-1})=\frac{\int h_t f(h_t|Y_{t-1})f(Y_{t-1})dh_t}{\int\int f(Y_{t-1},u_{t-1},h_{t-1}) dh_{t-1}du_{t-1}}\\
&=\frac{\int h_t \Big\{\int f(h_t|h_{t-1})\big[\int f(Y_{t-1},u_{t-1},h_{t-1})du_{t-1}\big]dh_{t-1}\Big\}dh_t}{\int\int f(Y_{t-1},u_{t-1},h_{t-1}) du_{t-1}dh_{t-1}}\\
&\simeq \frac{\frac{e-d}{2}\sum_j^{n_h} h_jw_{hj} \sum_{j'}^{n_h} F_{hjj'}w_{hj'}\sum_{i}^{n_u} l_{t-1ij'}}{\sum_{i}^{n_u}\sum_j^{n_h} l_{t-1,ij}}\\
\hat{u}_t(Y_{t-1}) &\simeq \frac{\frac{b-a}{2}\sum_i^{n_u} u_iw_{ui} \sum_{i'}^{n_u} F_{uii'}w_{ui'}\sum_{j}^{n_h} l_{t-1,i'j}}{\sum_{i}^{n_u}\sum_j^{n_h} l_{t-1,ij}}
\end{align*}
where $F_{uii'}$ and $F_{hjj'}$ denote the generic elements of the transition matrices $F_{u}$ and $F_{h}$ respectively.

\section{Application to tribal Art prices}
In this section we illustrate the application of our model to the first database of ethnic artworks. The responses are the logged prices for $28$ semesters for the overall sample size of 13955 items. We take the fixed effects hedonic specification (FE) as the benchmark model. The covariates are selected through stepwise (forward and backward) techniques and then combining parsimony with Art Economics arguments. Table~\ref{tab:results} reports the parameter estimates for the three models: fixed effects (FE), autoregressive random effects (ARE) and stochastic volatility with autoregressive random effects (SVARE), fitted on the same data set with the same set of covariates. The asymptotic standard errors for the FE and SVARE models are derived from the Hessian matrix of the likelihood functions. The robust standard error for the ARE model are derived by means of the wild bootstrap for multilevel models introduced in \cite{Mod15}.
\par
In order to assess whether the specifications proposed manage to model satisfactorily the time dynamics and the heterogeneity observed, we have implemented a series of diagnostic tests. Table~\ref{tab:AIC} reports information on the goodness of fit of the models whereas in Table~\ref{tab:shapiro} we present the results of some diagnostic tests and indicators on the residuals. In particular, the first row of Table~\ref{tab:shapiro} shows the $p$-values for the Shapiro-Wilk test for normality of the residuals of the three models; the \texttt{shapiro.test} function in \texttt{R} limits the sample size to 5000. The results presented are the median $p$-values over 20000 random subsamples of size 5000 drawn from the original sample. The last two rows show the indexes of skewness and kurtosis $b_1$ and $b_2$ as in \cite{Joa98} computed on level-1 residuals. We derive the standardized residuals for the ARE model from the BLUP of the random effects whereas for the SVARE model we use the smoothed values for both the random effects (Eq.~\ref{ut}) and the volatility (Eq.~\ref{ht}).
\par
As concerns the time dynamics we assess the adequateness of the models by computing the sample global and partial autocorrelation functions over time varying quantities such as level-2 residuals. Moreover, we use the metric entropy measure $S_k$ defined as
\begin{equation}\label{Srho}
S_k=\frac{1}{2}\int\int\left[ \{f_{(X_{t},X_{t+k})}(x_{1},x_{2})\}^{1/2}-\{f_{X_{t}}(x_{1})f_{X_{t+k}}(x_{2})\}^{1/2}\right] ^{2}dx_{1}dx_{2}
\end{equation}
where $f_{X_{t}}$ and $f_{(X_{t},X_{t+k})}$ denote the probability density function of $X_{t}$ and of the vector $ (X_{t},X_{t+k})$ respectively. The measure is a particular member of the family of relative entropies, which includes as a special case non-metric entropies often referred to as Shannon or Kullback--Leibler divergence. It can be interpreted as a nonlinear autocorrelation function and possesses many desirable properties. We use $S_k$ as in \cite{Gia15} to test for nonlinear serial dependence and as in \cite{Gra04} to test for serial independence (see the supplementary material for more details). The tests are implemented in the \texttt{R} package \texttt{tseriesEntropy} \cite{tseriesEntropy}. In the spirit of time series analysis, if the specification is appropriate then the residuals behave as a white noise process and diagnostic tests can suggest directions to improve the existing model.
\par
Finally, Table~\ref{tab:pred} summarizes and compares the prediction/forecasting capability of the three models under scrutiny. The aggregate measures of prediction error are the Mean Absolute (Prediction) Error (MAE) and the Root Mean Square (Prediction) Error (RMSE):
\begin{equation}\label{eq:pred}
\text{MAE}=\frac{1}{n_{T+1}}\sum_{i=1}^{n_t}\left|y_{i,T+1}-\hat{y}_{i,T+1}\right| \,;\, \text{RMSE}=\sqrt{\frac{1}{n_{T+1}}\sum_{i=1}^{n_{T+1}}\big(y_{i,T+1}-\hat{y}_{i,T+1}\big)^2}
\end{equation}
The first two rows of Table~\ref{tab:pred} report the prediction error over 100 (out of sample) items within the time span 1998-2011,  and the last two rows of the table show the forecasting performance over all the 73 observations of semester 2011-2. Such observations have not been included in the model so that the measures reflect a genuine forecasting performance.

\begin{table}
\centering \scriptsize
\caption{\label{tab:results}
Parameter estimates for models FE, ARE and SVARE with standard errors in parentheses.
For each categorical variable the baseline category is indicated. The complete set of estimates is available in the supplementary material}
\begin{tabular}{lrrr}
  \hline
                                        & FE & ARE & SVARE \\
  \hline
  $\sigma^2$                      &  0.225 (0.006) &  0.226 (0.004) & -              \\
  $\sigma^2_{\eta}$               & -              &  0.022 (0.015) &  0.021 (0.000) \\
  $\rho$                          & -              &  0.837 (0.140) &  0.848 (0.025) \\
  $\alpha$                        & -              &  -             & -0.142 (0.012) \\
  $\delta$                        & -              &  -             &  0.931 (0.035) \\
  $\sigma^2_{\nu}$                & -              &  -             &  0.158 (0.001) \\
  CABS (Yes vs No)                &  0.103 (0.013) &  0.102 (0.015) &  0.093 (0.012) \\
  CABC (Yes vs No)                &  0.138 (0.010) &  0.138 (0.011) &  0.127 (0.010) \\
  CAES (Yes vs No)                &  0.078 (0.015) &  0.078 (0.016) &  0.081 (0.013) \\
  Christie's (vs Sotheby's)       & -0.109 (0.011) & -0.110 (0.013) & -0.153 (0.011) \\
  Paris (vs New York)             & -0.126 (0.013) & -0.124 (0.015) & -0.076 (0.012) \\
    \multicolumn{4}{l}{Illustration: baseline Absent}\\
  - Miscellaneous col.            &  0.452 (0.021) &  0.453 (0.031)&  0.441 (0.020) \\
  - Col. cover                    &  1.494 (0.076) &  1.494 (0.104)&  1.395 (0.075) \\
  - Col. half page                &  0.996 (0.025) &  0.995 (0.034)&  0.864 (0.024) \\
  - Col. full page                &  1.148 (0.026) &  1.149 (0.034)&  1.030 (0.026) \\
  - More than one col.            &  1.376 (0.028) &  1.377 (0.037)&  1.273 (0.027) \\
  - Col. quarter page             &  0.823 (0.021) &  0.823 (0.029)&  0.701 (0.020) \\
  - Miscellaneous b/w             &  0.500 (0.047) &  0.500 (0.040)&  0.411 (0.038) \\
  - b/w half page                 &  0.647 (0.061) &  0.647 (0.057)&  0.553 (0.047) \\
  - b/w full page                 &  1.024 (0.239) &  1.025 (0.290)&  0.829 (0.190) \\
  - b/w quarter page              &  0.383 (0.029) &  0.383 (0.033)&  0.302 (0.025) \\
   \multicolumn{4}{l}{Description on the catalogue: baseline Absent}\\
  - Short visual descr.           & -0.099 (0.037) & -0.100 (0.042) & -0.138 (0.035) \\
  - Visual descr.                 &  0.078 (0.038) &  0.076 (0.044) &  0.036 (0.035) \\
  - Broad visual descr.           &  0.325 (0.041) &  0.325 (0.047) &  0.274 (0.038) \\
  - Critical descr.               &  0.319 (0.041) &  0.319 (0.049) &  0.266 (0.038) \\
  - Broad critical descr.         &  0.719 (0.045) &  0.718 (0.055) &  0.640 (0.042) \\
  \multicolumn{4}{l}{Type of object: baseline Furniture}\\
  - Masks                         &  0.121 (0.025) &  0.121 (0.025)&  0.111 (0.023) \\
  - Religious objects             &  0.047 (0.027) &  0.046 (0.030)&  0.039 (0.025) \\
  - Ornaments                     & -0.161 (0.026) & -0.161 (0.030)& -0.111 (0.024) \\
  - Sculptures                    &  0.062 (0.023) &  0.061 (0.025)&  0.062 (0.021) \\
  - Musical instruments           & -0.098 (0.036) & -0.098 (0.044)& -0.061 (0.033) \\
  - Tools                         & -0.088 (0.024) & -0.089 (0.027)& -0.075 (0.022) \\
  - Clothing                      & -0.077 (0.037) & -0.078 (0.040)& -0.075 (0.032) \\
  - Weapons                       & -0.131 (0.030) & -0.131 (0.033)& -0.095 (0.028) \\
\multicolumn{4}{l}{Continent: baseline Africa}\\
  - America                       &  0.175 (0.013) &   0.174 (0.016)&  0.162 (0.013) \\
  - Oceania                       &  0.167 (0.012) &   0.167 (0.014)&  0.147 (0.011) \\
   \multicolumn{4}{l}{Patina: baseline Not indicated}\\
  - Pejorative                    &  0.210 (0.046) &  0.209 (0.044) &  0.167 (0.046) \\
  - Appreciative                  &  0.124 (0.012) &  0.124 (0.014) &  0.104 (0.012) \\
  \multicolumn{4}{l}{Historicization: baseline Absent}\\
  - Museum certification          &  0.045 (0.017) &  0.045 (0.021) &  0.021 (0.016) \\
  - Relevant museum certification &  0.065 (0.016) &  0.064 (0.018) &  0.045 (0.015) \\
  - Simple certification          &  0.053 (0.011) &  0.053 (0.013) &  0.061 (0.010) \\
   \hline
\end{tabular}
\end{table}
\begin{table}
\caption{Loglikelihood, number of parameters and Information criteria for the hedonic regression (FE), ARE and SVARE models.}\label{tab:AIC}
\begin{tabular}{llll}
\toprule
 & FE & ARE & SVARE \\
\midrule
loglik & -9405.833 & -9472.96  & -8785.624 \\
n. par & 70        & 45        &    47     \\
  AIC  &  18952    & 19036     & 17665     \\
  BIC  & 19480     & 19375     & 18020     \\
\bottomrule
\end{tabular}
\end{table}
\subsection{FE and ARE models}
  The parameter estimates of the FE and ARE models are very similar (first two columns of Table~\ref{tab:results}), still, there are important differences: first, the ARE fit is more parsimonious and results in a smaller BIC and AIC (see Table~\ref{tab:AIC}); also, it provides a decomposition of the total variability of the response in between-time and within-time variability. Furthermore, in the FE model the time dynamics is modelled through 28 dummy variables while in the ARE model the time dynamics is fully captured through the AR(1) specification with the two parameters, $\rho$ and $\sigma_{\eta}$, see Figure 1 and 2 of the Supplementary Material (SM from now on). The same tests performed on level-2 residuals of the ARE model show no structure (see Figure 3 and 4 of the SM). Finally, the ARE model provides a superior one-step-ahead forecasting of the price whereas the prediction performance is the same as that of the FE model (see Table~\ref{tab:pred}).
\begin{table}
\caption{$p$-values of the Shapiro-Wilk test and indexes of skewness and kurtosis for the FE, ARE and SVARE (level-1) residuals.}\label{tab:shapiro}
\begin{tabular}{lrrr}
\toprule
 & FE & \multicolumn{1}{c}{ARE }& \multicolumn{1}{c}{SVARE }\\
\midrule
  Shapiro-Wilk $p$-value  & $<$2e-17 & 2e-16   & 3e-07  \\
  Skewness                & -0.1683  &-0.1727         & 0.0221         \\
  Kurtosis                &  1.0553  & 1.0630         & 0.4568         \\
\bottomrule
\end{tabular}
\end{table}
The Shapiro-Wilk test (see Table~\ref{tab:shapiro}) points to a deviation from normality in level-1 residuals of the ARE model (whereas it does not reject the assumption of normality for level-2 residuals). As discussed above, this is consistent with the findings in literature and might be due to heteroscedasticity. In fact, similarly to other assets, level-1 residuals show a leptokurtic behaviour as shown in Figure~\ref{fig:res} and by looking at the kurtosis index in Table~\ref{tab:shapiro}. Furthermore, we reject the assumption of homogeneity of the variance across time points, tested through a non-parametric version of the \citet{Levene} rank-based test \citep{Kruskal}.
\begin{figure}
\centering
\includegraphics[scale=.35]{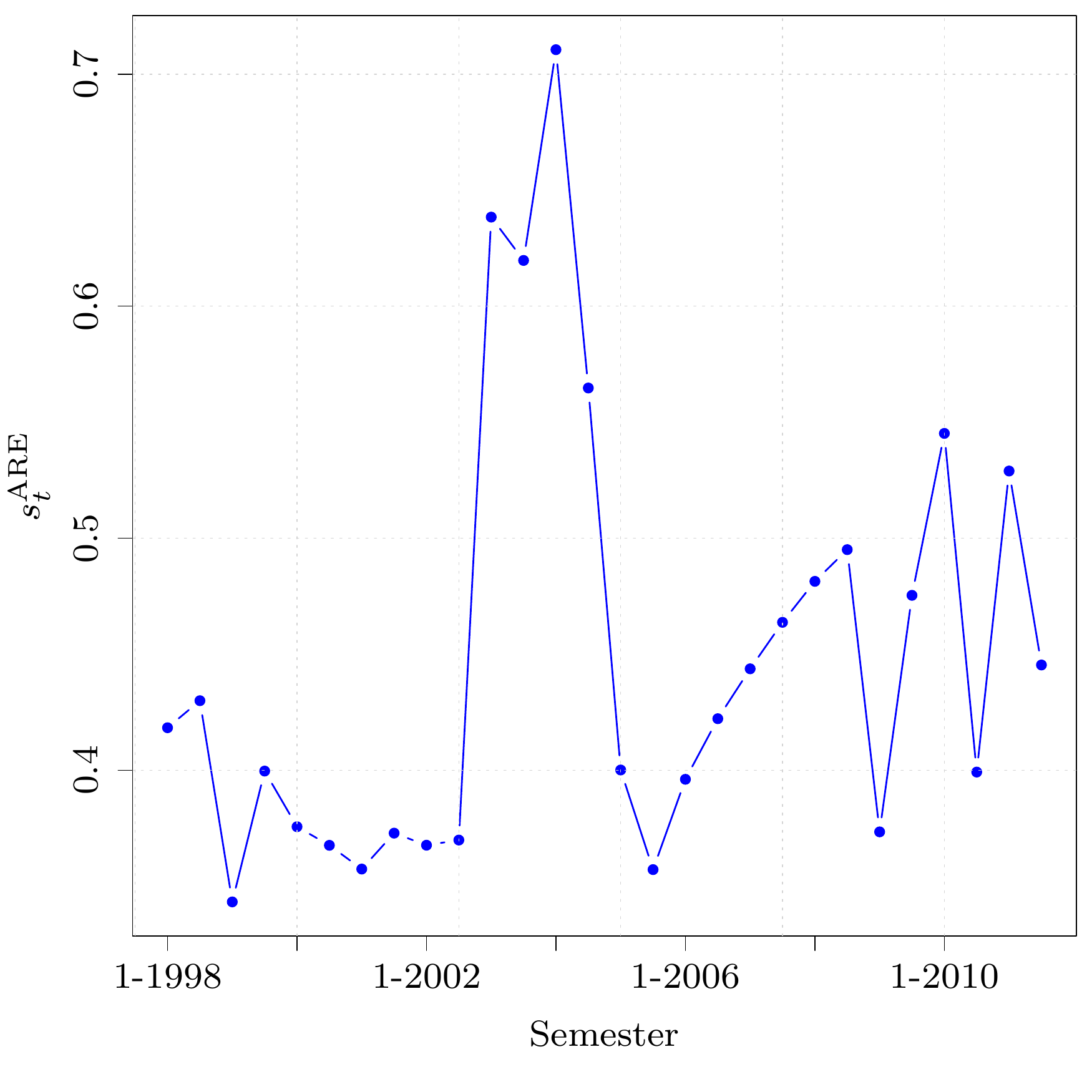}
\includegraphics[scale=.35]{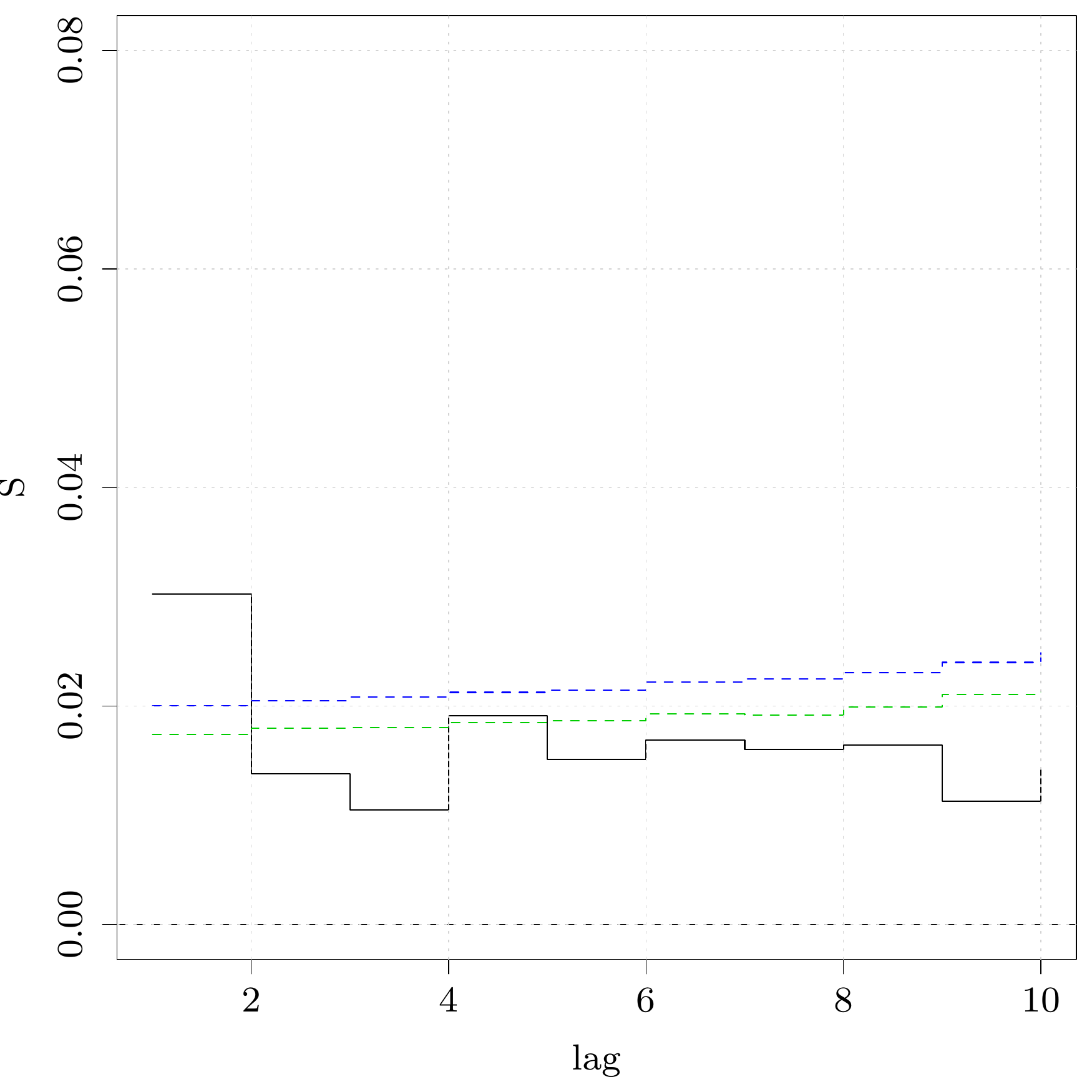}
\caption{Standard deviations of ARE level-1 residuals $s^{\text{ARE}}_t$: time plot (left) and entropy measure of dependence (right). The confidence bands correspond to the null hypothesis of serial independence at levels 90\% and 95\% up to 10 lags/semesters.}\label{fig:sd}
\end{figure}

\begin{figure}
\includegraphics[scale=.35]{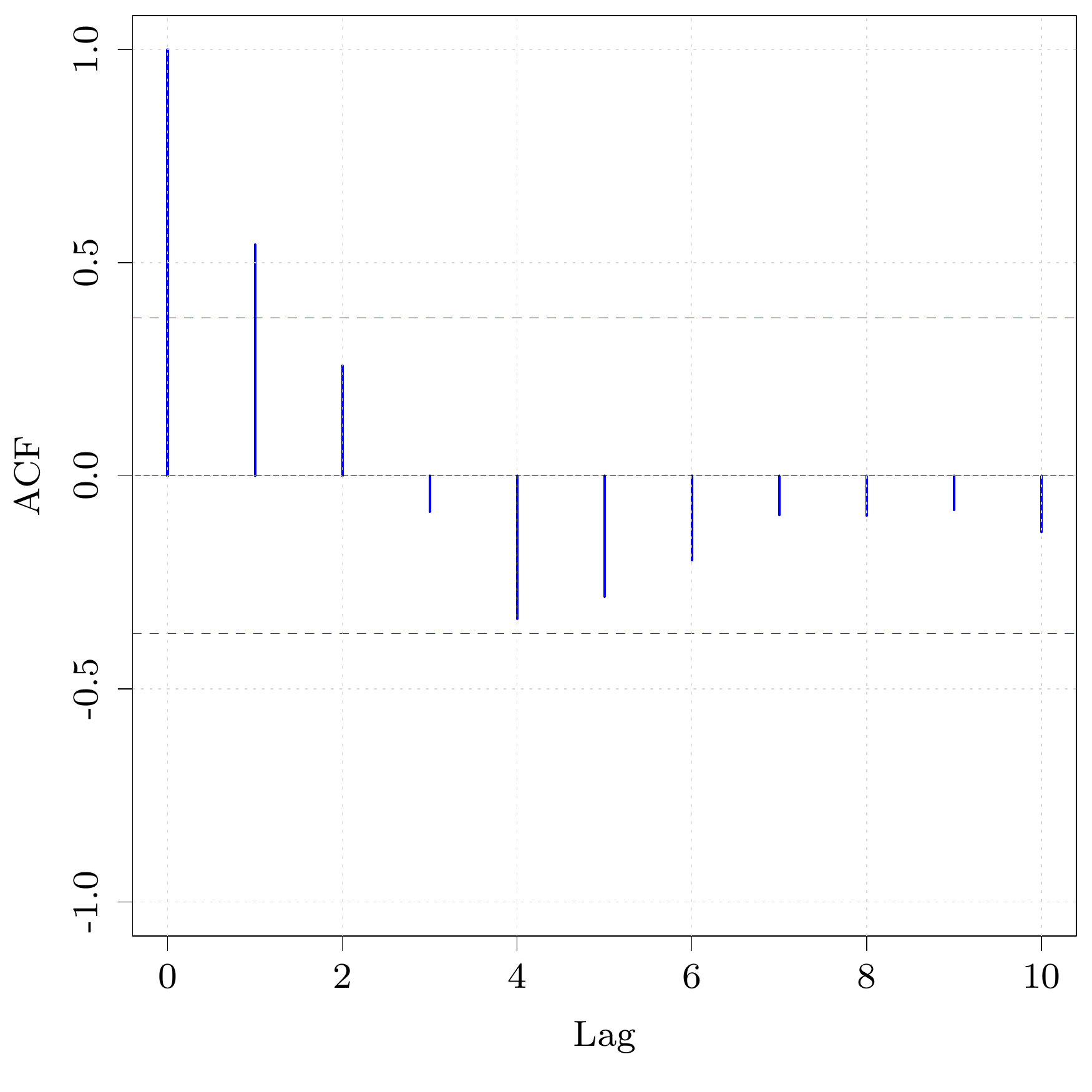}
\includegraphics[scale=.35]{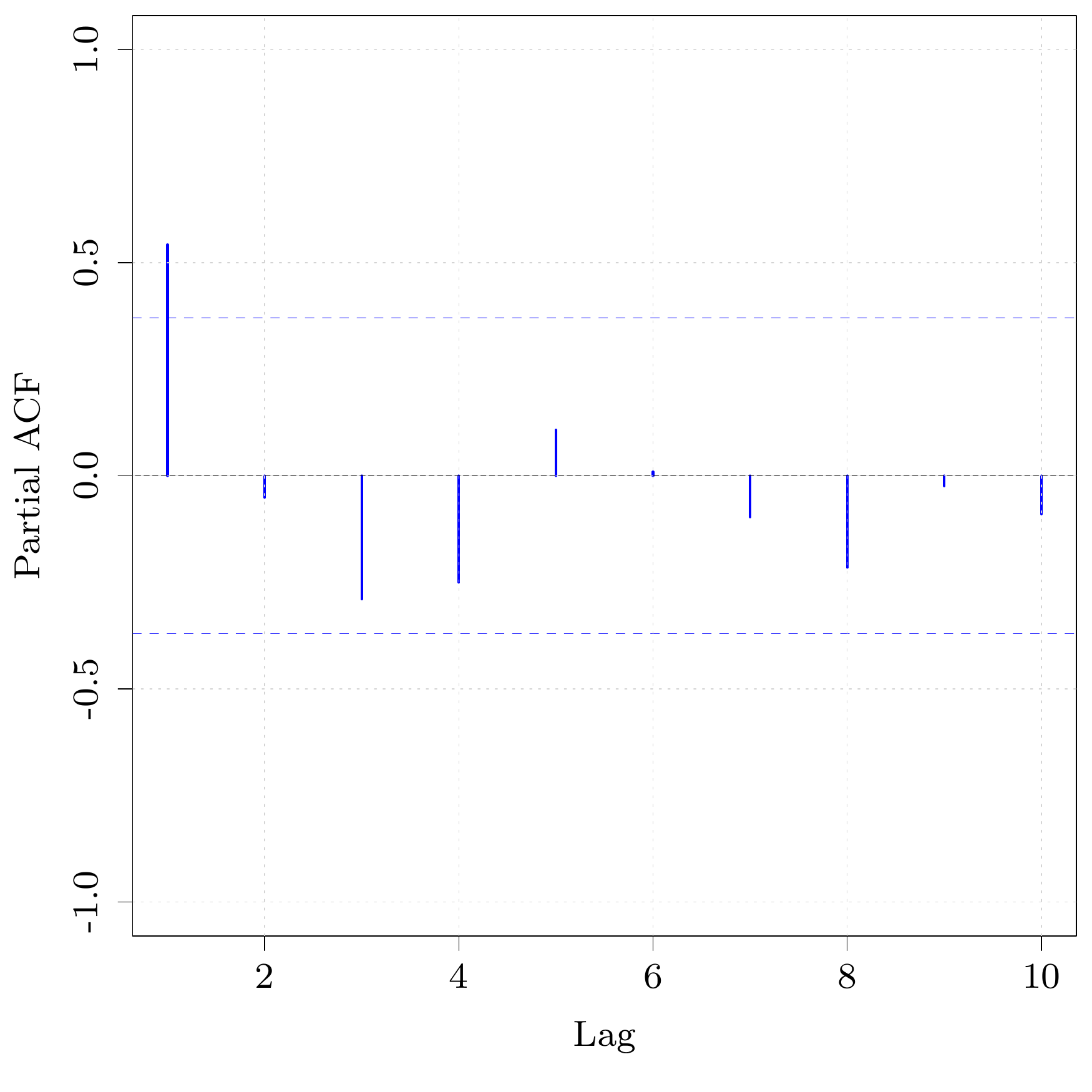}
\caption{Global (left) and partial (right) empirical autocorrelation functions of standard deviations of ARE level-1 residuals.}\label{fig:acfsd}
\end{figure}

\begin{figure}
\includegraphics[scale=.35]{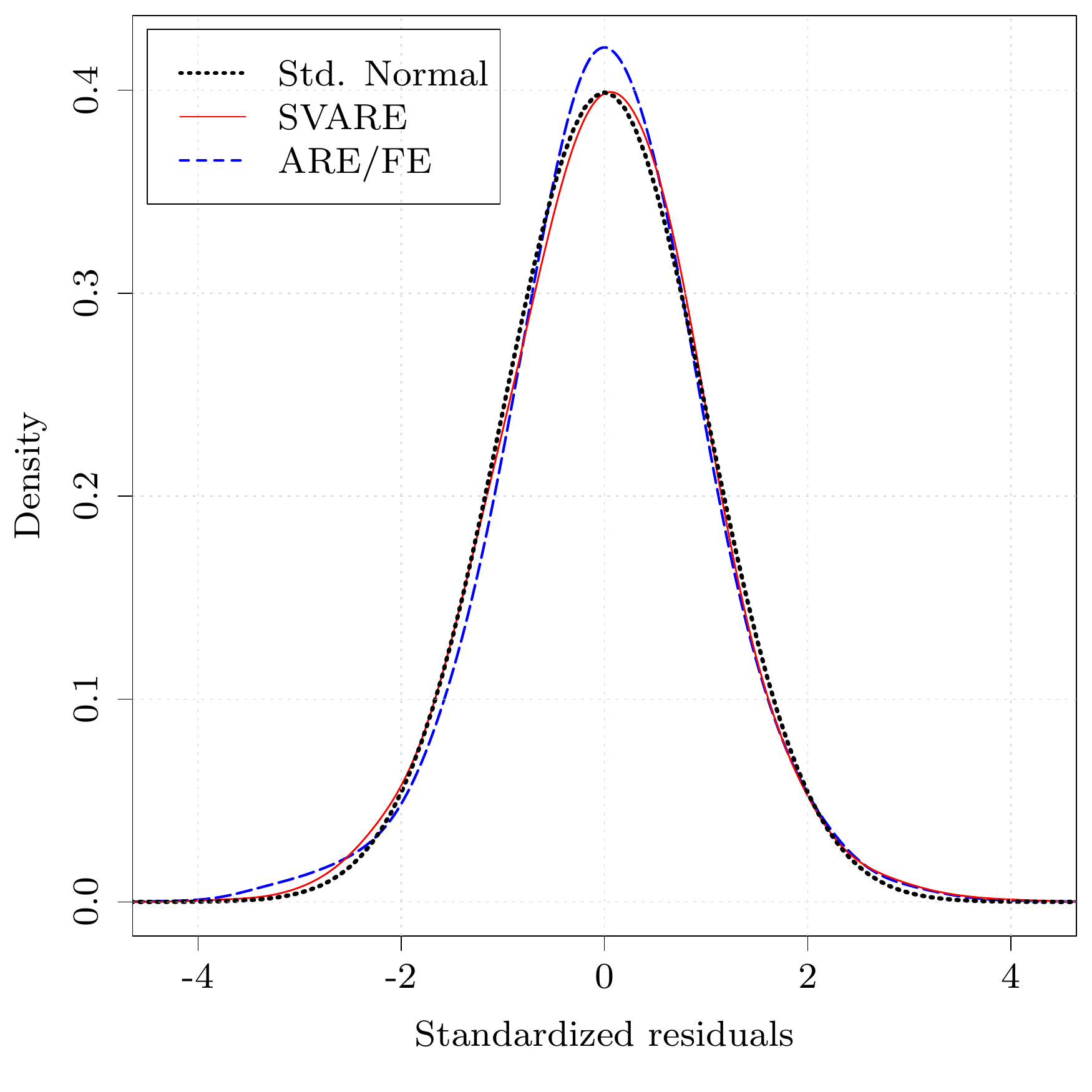}
\includegraphics[scale=.35]{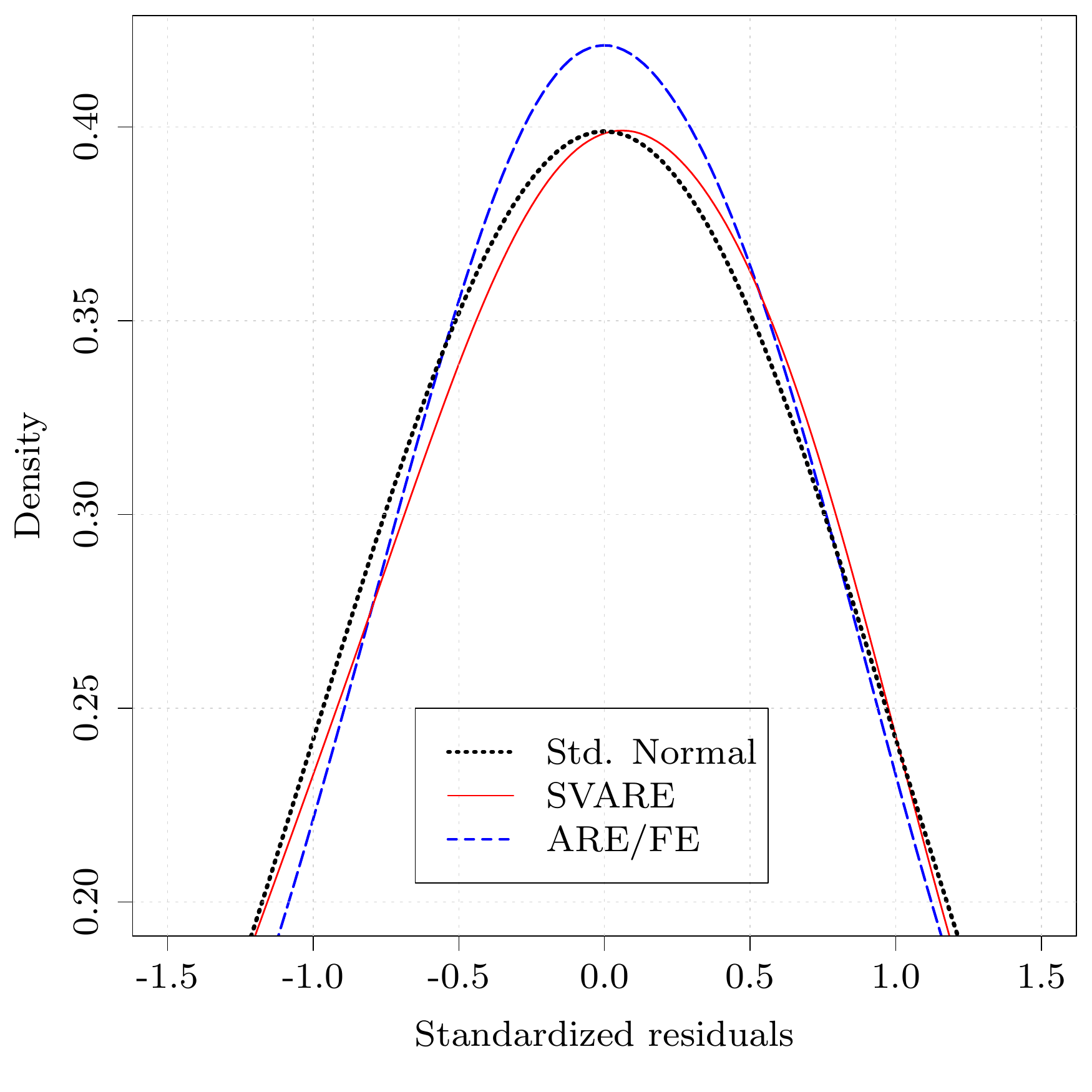}
\caption{Density of level-1 residuals. The right panel is a zoomed detail of the peak that highlights the kurtosis.}\label{fig:res}
\end{figure}
The plot of $s^{\text{ARE}}_t$, the standard deviations of level-1 residuals $\hat\epsilon_{it}$ in Figure~\ref{fig:sd}(left) provides a visual evidence of volatility patterns. The entropy measure $S_k$ shown in Figure~\ref{fig:sd}(right) confirms the presence of a linear serial dependence (the test for nonlinearity does not reject, see Figure 6 of the supplementary material) and the correlograms of Figure~\ref{fig:acfsd} indicate a AR(1)-type dependence structure for the volatility. In the following subsection we account for the observed heterogeneity by fitting the multilevel model with autoregressive random effects and stochastic volatility (SVARE).

\subsection{SVARE model}
A first important result is that the parameter estimates differ to some extent from those of the FE/ARE models, see Table~\ref{tab:results}. Even if in most cases the significance of the parameters does not change, SVARE estimates account properly for the volatility and reflect more closely the impact of the covariates on artwork prices. The estimate of the volatility parameter $\hat\delta=0.931$ agrees with those of models for financial time series reported in literature and indicates a non-negligible volatility persistence. Indeed, the goodness of fit of the SVARE model increases considerably as witnessed by the information criteria shown in Table~\ref{tab:AIC}. Note also that the standard errors of the estimates are almost uniformly the smallest among the three models.
\par
Also for the SVARE model, the Shapiro-Wilk test of normality of level-1 residuals rejects the null hypothesis (Table~\ref{tab:shapiro}). Nevertheless, the leptokurtic behaviour of residuals is considerably reduced with respect to both the ARE and FE models. This is shown in Table~\ref{tab:shapiro} (the skewness disappears and the kurtosis is more than halved) and in Figure~\ref{fig:res} where we show the densities of level-1 residuals for the ARE and SVARE models. Note the agreement of SVARE residuals with the standard Normal density (dotted in the figure). See also Figure 11 of the SM for a normal qqplot of level-1 residuals of the two models. As in the FE/ARE case, we compute the diagnostic tests of dependence on level-2 residuals $\hat{\eta}_t$ and $\hat{\nu}_t$. Both the correlograms and the entropy measure $S_k$ indicate the absence of any dependence structure (see Figures 7 - 10 of the Supplementary Material) so that we may argue that the SVARE specification manages to capture the volatility dynamics.
\begin{table}
\caption{\label{tab:pred} Prediction/forecasting performance of the three models over 100 out-of-sample units within the time span 1998-2011 (rows 1-2) and over 73 units of the out-of-sample semester, 2011-2 (rows 3-4).}
  \begin{tabular}{llccc}
\toprule
    &&  FE& ARE & SVARE\\
\midrule
Prediction & MAE  & 0.288 & 0.288 & 0.278 \\
           & RMSE & 0.355 & 0.355 & 0.344 \\
    \hline
Forecast & MAE  & 0.534 & 0.376 & 0.355 \\
         & RMSE & 0.673 & 0.512 & 0.479 \\
\bottomrule
    \end{tabular}
\end{table}
Finally, from Table~\ref{tab:pred} it emerges clearly that the SVARE specification performs best among competitors in terms of both prediction and forecasting.
\par
Note that the choice of the number of quadrature points $n_u$ and $n_h$ is crucial to the estimation of the SVARE model. We first set $n_u =21$ and $n_h = 41$ according to the rule given in Section~\ref{sec:lik}, then we increased them up to $71$. As for the autoregressive part of the model, we found that the parameters' estimates are quite robust except for $\rho$ whose estimate stabilizes for $n_u \geq 51$. The estimates of the Stochastic Volatility component are more sensitive to the number of quadrature points so that we set $n_u = n_h=61$. This yields sensible results even for high values of $\delta$, that is, in those cases where the integrand is very peaked and requires a large number of quadrature points to be well approximated.
\section{Conclusions}
\begin{figure}
\includegraphics[scale=.35]{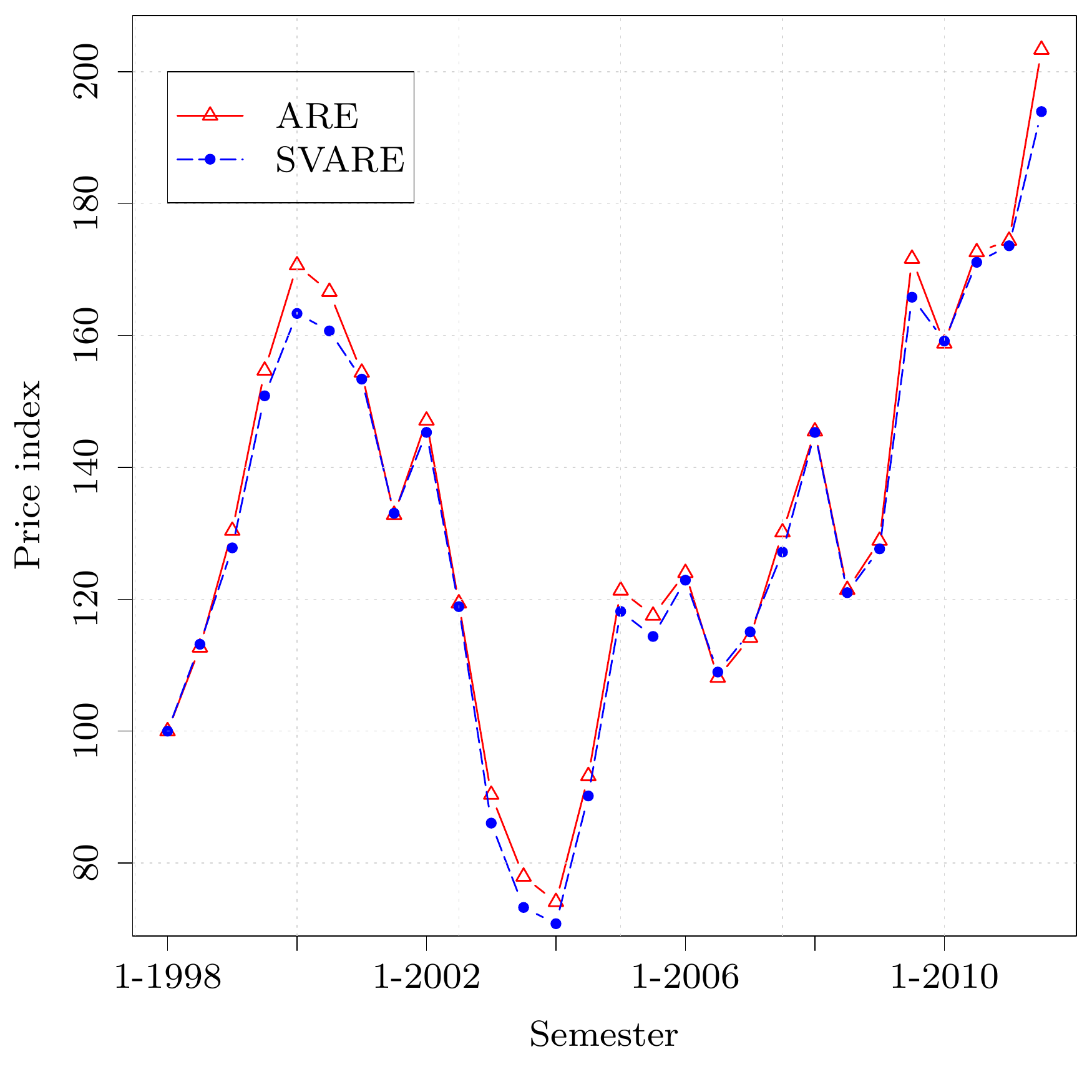}
\includegraphics[scale=.35]{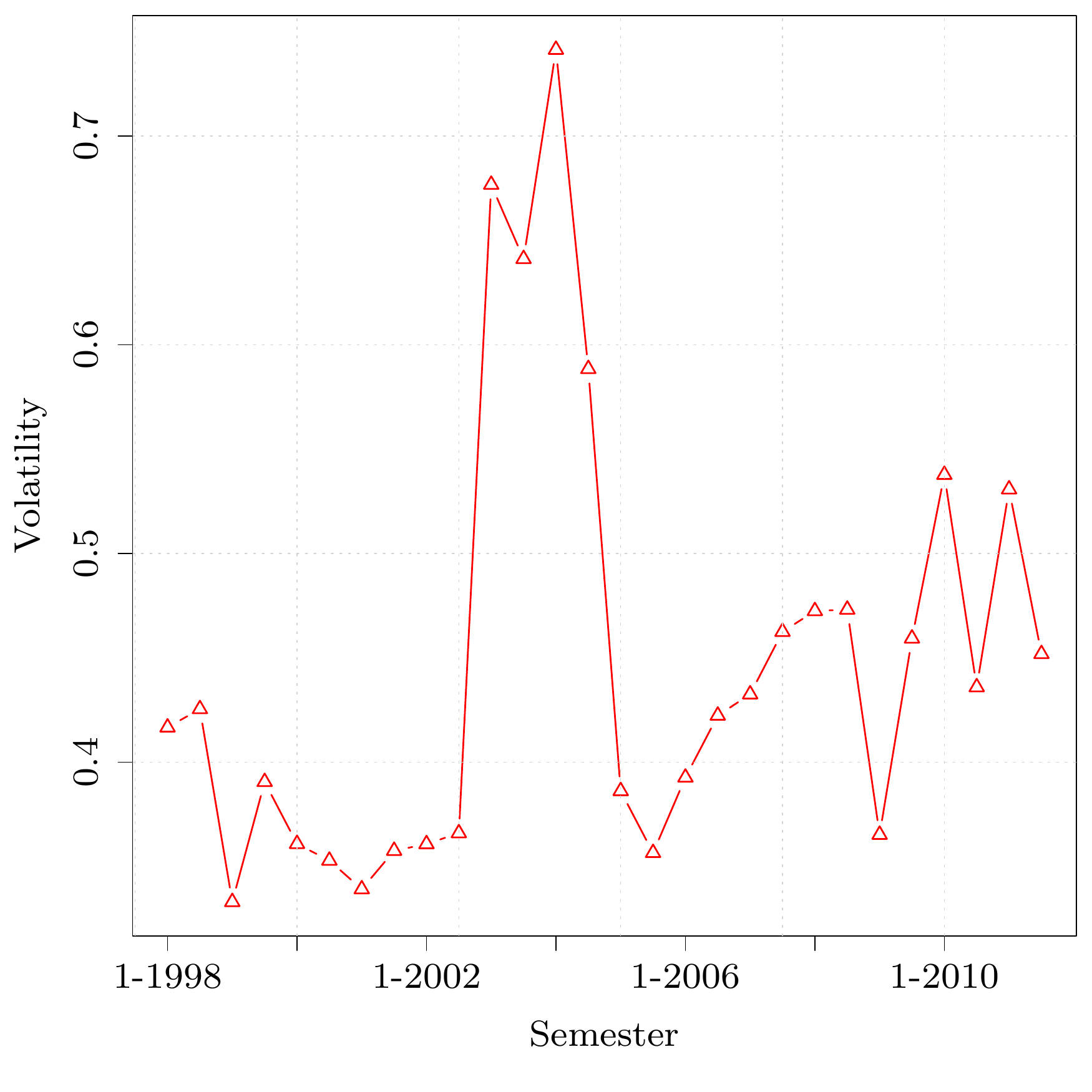}
\caption{Price index (with base ``semester 1-1998'') of the art market for the ARE and SVARE models (left panel). Plot of volatility of the SVARE model (right panel).}\label{fig:index}
\end{figure}

A great added value of the SVARE specification for repeated cross sections is that it encompasses in a unique framework the trends in the mean and in the volatility of artwork prices. The left panel of Figure~\ref{fig:index} compares the resulting biannual price indexes obtained through the ARE and SVARE fits. They are computed with fixed base in semester $b=$``1-1998'' as \[I_t=\frac{e^{\hat{\beta}_{0t}}}{e^{\hat{\beta}_{0b}}}\times 100,\]
where $\hat{\beta}_{0t}$ are the BLUP values $\hat{\beta}_0+\hat{u}_t$. Note that the indexes are similar, nevertheless, the SVARE model provides the predicted volatility values $\hat h_t$ of Eq.~\ref{ht} (right panel of Figure~\ref{fig:index}). This is important additional information on the predictability of the prices that complements that of the index and can be exploited by art market stakeholders for an informed decision making.


\section*{Acknowledgements}
We would like to thank Guido Candela, Antonello Scorcu, Massimiliano Castellani and Pierpaolo Pattitoni for useful discussions. Silvia Cagnone and Lucia Modugno acknowledge the financial support from the grant RBFR12SHVV funded by the Italian Government (FIRB project ``Mixture and latent variable models for causal inference and analysis of socio-economic data").

\bibliographystyle{chicago}
\bibliography{biblioSV}
\end{document}